%% file: main.tex
\documentclass[journal]{vgtc} 

\usepackage{cite}
\usepackage[bottom]{footmisc}
\usepackage{mathptmx}
\usepackage{graphicx}
\usepackage{siunitx}
\usepackage{times}
\usepackage[T1]{fontenc}
\usepackage[latin1]{inputenc}
\usepackage{enumitem}
\usepackage[table]{xcolor} 

\usepackage{dirtytalk}
\usepackage{soul}
\usepackage{mdframed}
\usepackage{xcolor}
\definecolor{CX}{rgb}{0, 0, 100}
\definecolor{TC}{rgb}{100,0,0}

\usepackage[bookmarks,backref=true,linkcolor=black]{hyperref} 
\hypersetup{
 pdfauthor = {},
 pdftitle = {},
 pdfsubject = {},
 pdfkeywords = {},
 colorlinks=true,
 linkcolor= black,
 citecolor= black,
 pageanchor=true,
 urlcolor = black,
 plainpages = false,
 linktocpage
}

\onlineid{0} 
\vgtccategory{Research} 
\vgtcinsertpkg 

\title{Biased Average Position Estimates in Line and Bar Graphs: Underestimation, Overestimation, and Perceptual Pull}

\author{Cindy Xiong, Cristina R. Ceja, Casimir J.H. Ludwig, and Steven Franconeri}
\authorfooter{ 
\item
Cindy Xiong is with Northwestern University. E-mail: cxiong@u.northwestern.edu.
\item
Cristina R. Ceja is with Northwestern University. E-mail: crceja@u.northwestern.edu.
\item
Casimir J.H. Ludwig is with University of Bristol. Email: c.ludwig@bristol.ac.uk
\item
Steven Franconeri is with Northwestern University. E-mail: franconeri@northwestern.edu.
}

\shortauthortitle{Xiong \MakeLowercase{\textit{et al.}}: Biased Average Position Estimates in Line and Bar Graphs: Underestimation, Overestimation, and Perceptual Pull}

\abstract{In visual depictions of data, position (i.e., the vertical height of a line or a bar) is believed to be the most precise way to encode information compared to other encodings (e.g., hue). Not only are other encodings less precise than position, but they can also be prone to systematic biases (e.g., color category boundaries can distort perceived differences between hues). By comparison, position's high level of precision may seem to protect it from such biases. In contrast, across three empirical studies, we show that while position may be a precise form of data encoding, it can also produce systematic biases in how values are visually encoded, at least for reports of average position across a short delay. In displays with a single line or a single set of bars, reports of average positions were significantly biased, such that line positions were underestimated and bar positions were overestimated. In displays with multiple data series (i.e., multiple lines and/or sets of bars), this systematic bias still persisted. We also observed an effect of ``perceptual pull'', where the average position estimate for each series was `pulled' toward the other. These findings suggest that, although position may still be the most precise form of visual data encoding, it can also be systematically biased.} 


\teaser{
 \centering
 \includegraphics[width=16.75cm, height=2.22cm]{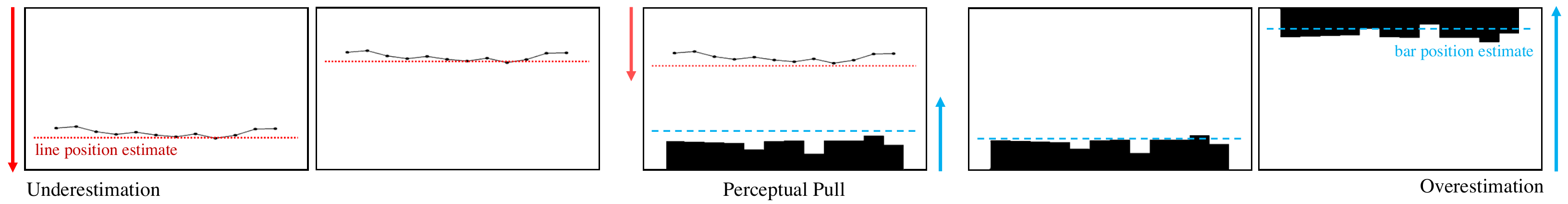}
 \caption{Reproductions of average line and bar positions are systematically biased, such that people tend to underestimate average line positions (toward the bottom of the display) and overestimate average bar positions (toward the top of the display). This bias appears in displays containing one data series (e.g., two outermost images on each side: a visualization containing only one line or only one set of bars). When a display consists of two data series (e.g., middle image: a visualization containing two lines, two bars, or a combination of the two), an additional effect of ``perceptual pull'' occurs: perception of average line and bar positions are pulled toward each other. This pull is shown to diminish or exaggerate the effect of underestimation and/or overestimation depending on the spatial arrangement of the data series representations.}
 \label{fig:teaser}
}

\begin{document}
\maketitle
\section{Introduction}
``Seeing is believing'' implies that vision delivers reality. Yet, experiments in perceptual psychology reveal that perception of a variety of visual information can be systematically influenced by recent history and context. Staring at downwardly moving dots can cause a `waterfall illusion' where people may perceive subsequent, static dots as moving in the opposite direction \cite{wade:1994:SelectiveHistory}. Similarly, a circle will appear smaller in size in the context of larger, concentric circles than when surrounded by smaller, concentric circles, as in the Ebbinghaus illusion \cite{roberts:2005:Ebbinghaus}. 

A small set of systematic biases based on history and context is already on the radar of visualization designers, such as how hue categories can bias the perception and memory of colors, or how background colors can strongly alter the perception of foreground colors \cite{ware:2012:InfoVisPerDesign}. But we generally assume that data visualizations are otherwise perceived in an unbiased (albeit, potentially noisy) manner, particularly for more precise visual data encodings such as position \cite{cm:1984:Graph}. 

The present experiments will show that, despite the high precision of position, perception of data from positional encodings can also be biased in systematic ways. Specifically, depending on the visualization, these data values can be significantly underestimated, overestimated, or `pulled' toward other irrelevant position values present in the same graph. 

\begin{figure*}[htb]
 \includegraphics[width=18.1cm]{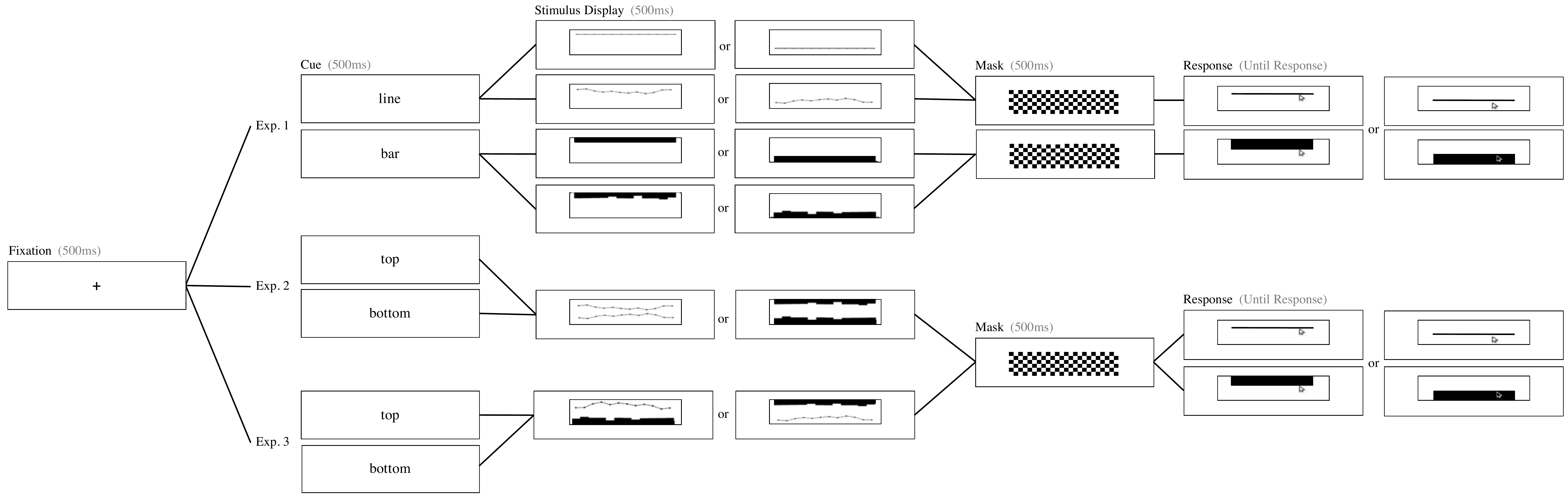}
 \caption{Experimental procedure and design for Experiment 1, 2, and 3. Display times shown in grey.}
 \label{fig:expTrial}
\end{figure*}

\section{Perceptual Biases}
The perception of visual magnitudes can be biased across multiple feature dimensions. Orientation estimates for a line can be either repulsed or attracted by the orientations of nearby objects, depending on the parameters of the display \cite{parkes:2001:Compulsory}. Perceived brightness can be affected by background brightness
\cite{ware:2012:InfoVisPerDesign}. Categorical boundaries between hues can exaggerate differences between those that straddle a boundary, compared to hues that do not \cite{bornstein:1984:Discrimination}. 

Similar patterns of categorical bias are found for object sizes once a set of object size categories have been learned \cite{kosslyn:1977:Category}. When a short ellipse is in the context of a taller ellipse, the short ellipse is perceived as even shorter and the taller ellipse as even taller \cite{sweeny:2011:Simultaneous}. After adapting to a vertically tall ellipse that is no longer present, the perceived height of a subsequent circle will appear vertically shorter \cite{kohler:1944:Figural}. Memory for the size of a single object in a crowd of other objects can be biased toward the average size of all objects \cite{ariely:2001:Statistical, chong:2003:Representation, brady:2011:Hierarchical}. 

Most importantly for the present study, even the perceived positions of objects can be biased, especially when positional changes break a categorical or relational boundary. Positions are encoded somewhat categorically (similar to color), so that changes to a category boundary are easier to detect, compared to equally distant changes that do not cross a category boundary \cite{kosslyn:1989:Categorical}. Viewers are more accurate at detecting change for a dot position when the dot is moved to the opposite side of a set of crossed gridlines, as opposed to moving to an equally distant position on the same side of the gridline \cite{kranjec:2014:Categorical}. When detecting changes to the spacing between pairs of circles, performance is higher when the changes affect the categorical position relations for the circles (`touching' to `not touching'), as opposed to when they do not (`not touching but close' to `not touching but far') \cite{kim:2012:Shapes, lovett:2017:Topological}. 

Most relevant for the present work is the susceptibility of position estimates to biases in memory, such that the position memory of a briefly presented object can be biased by context or nearby salient points in the display. Memory for the position of an object can be biased toward the average position of an associated group of objects \cite{alvarez:2008:Ensemble, lew:2015:Ensemble}. The recalled position of a dot inside a circle is pulled toward the center of one of four imaginary quadrants within the circle \cite{huttenlocher:1991:Categories}, suggesting that the nearest of these salient category boundaries pulled the position representation toward itself during recall. 

\section{Cognitive Biases}
Similar effects of irrelevant `background' values can bias cognitive processing, such as magnitude estimates at the level of verbal, numerical reports. In the `anchoring effect' \cite{tversky:1974:Judgment, furnham:2011:Literature}, uncertain target estimates can be strongly biased by other provided values, even if they are objectively irrelevant. For example, a population estimate of Nova Scotia would be biased by introducing an irrelevant value (e.g. ``Is the population of Nova Scotia more or less than 200,000?''), such that a larger primed value (e.g., 200,000) would lead to a larger population estimate, while a smaller primed value (e.g., 20,000) would lead to a smaller estimate \cite{jacowKahneman:1995:Anchoring}.

Such higher-level cognitive influences may also affect pattern perception in data visualizations. Socially-derived information signals (e.g., polls) can influence graph perception, such that other individuals' judgment of graphical information can bias how a single individual perceives and judges the same information \cite{hullmanAdarShah:2011:Socialonvisual}. The intensity of title-wording can cause graph viewers to overestimate or underestimate the slope of an associated, noisy scatterplot line, such that viewers who saw a high-intensity title recall a steeper slope than those who saw a low-intensity title \cite{newmanharozzoya:2018:Titlewordingtrendlines}. Previously viewed scatterplots can also influence viewer judgments of class separability in novel scatterplots, which could be interpreted similarly as an anchoring effect \cite{valdez:2018:Priming}. For example, priming with a clearly separable point cloud can bias perception of an ambiguous point cloud to appear more separable than if primed with a non-separable point cloud. 

\section{Contributions} 
The current work bridges vision science research and visualization research to test for perceptual biases with graphed data series.

Much of the previous literature investigating perceptual biases has focused on biases within visualizations in a higher-level, cognitive context --- how do elements (e.g., priming, titles, axes, etc) influence perception of relevant data? In the current study, we are interested in potential biases found in lower-level perception of data --- can people accurately perceive purely graphical information? Specifically, we focus on possible biases within average position estimations to better understand perception of distributed visual information in both simple and complex graphs. Position averaging is not only one of the most common and crucial tasks when interpreting visualized data, but also an area few have studied.

\section{Study Overview}
Across three experiments, we apply designs and methods from perceptual psychology to capture potential biases in average position perception of data series in visualized data.

We first explore how average positions of simple graphed data series are perceived in a single visualization to test how people perceive the average position of a line within a line graph, or a set of bars within a bar graph. We then explore how graphing multiple lines and/or sets of bars on the same graph may influence perception of the average position of a target line or target set of bars. 

Our findings show that people systematically perceive graphed data series in a biased manner\footnote{Note that any mention of ``underestimation'' or ``overestimation'' is in relation to the 140-pixel display frame, where values of $[0,140]$ map onto the bottom and top of the frame, respectively. For example, for a set of top (downward-pointing) bars with an average of 100 pixels, overestimation (100-140 pixels) in this display-based frame will actually reflect shorter bar lengths. We chose this naming convention to be more closely aligned with real-world scenarios where downward-pointing bars are observed, such as bars that depict negative values. See Figure~\ref{fig:teaser} for clarification.}, underestimating the average positions of lines and overestimating the average positions of bars in a graph. We also found that under- and over-estimation of the target line or set of bars, respectively, can be exaggerated or diminished by the presence of other such graphed data series. The average position estimates for lines or bars tend to gravitate toward the positions of the other lines and/or bars present on the same graph. We call this perceptual bias of irrelevant, graphed data series on relevant, targeted series ``perceptual pull''.

\begin{figure*}[htb]
 \includegraphics[width=18.1cm]{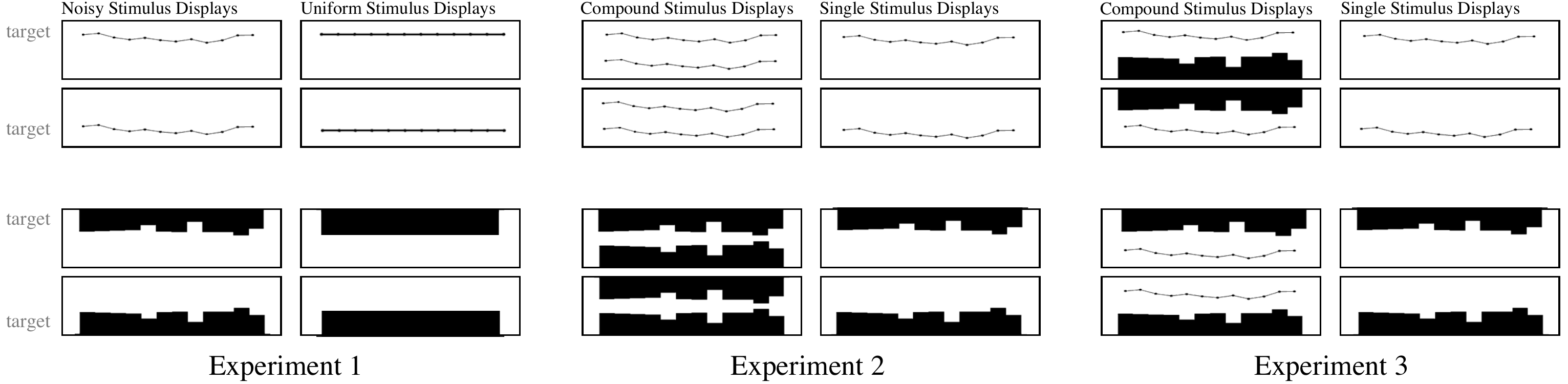}
 \caption{Design space for Experiment 1, 2, and 3. Experiment 1 investigates position perception bias in simple line and bar graphs for a single line or a single set of bars, respectively, which could be noisy or uniform. Experiment 2 examines this bias in complex graphs with either two lines or two sets of bars present in the same display, to be compared against displays with a simple single line or a single set of bars. Experiment 3 examines combined data series with one line and one set of bars present in the same display, compared to displays with a simple single line or a single set of bars. Each possible display type was generated with a data series centered around the three means shown in Figure {\ref{fig:meanValues}}, for lines and for bars, respectively.}
 \label{fig:designSpace}
\end{figure*}

\subsection{General Stimulus and Procedure}
All experimental stimuli were generated using MATLAB with the Psychophysics Toolbox \cite{brainard:1997:Psychophysics, kleiner:2007:Psychtoolbox, pelli:1997:Videotoolbox} driven by an Apple Mac Mini running OS 10.10.5. The monitor was 21-inch with a resolution of 1280 x 800 pixels and a refresh rate of 60 Hz. The average viewing distance was approximately 47 cm. All experimental materials and analysis are included in the supplementary materials.

As shown in Figure~\ref{fig:expTrial}, each stimulus display contained a graph enclosed in a display frame (9.6 x 2.8 inches; 538 x 140 pixels). All position estimations were recorded and analyzed relative to the 140 pixel display frame, where the bottom of the display frame is 0 and the top is 140. Graphs within the display frame varied in twelve types (see ``Stimulus Display'' in Figure~\ref{fig:expTrial}, or Figure \ref{fig:designSpace}). To balance the stimuli, we generated three mean values (high, medium, and low) for lines or bars located in the top half of the display, and three mean values (high, medium, and low) for lines or bars located in the bottom half.

\begin{figure}[b!]
 \includegraphics[width=3.4in]{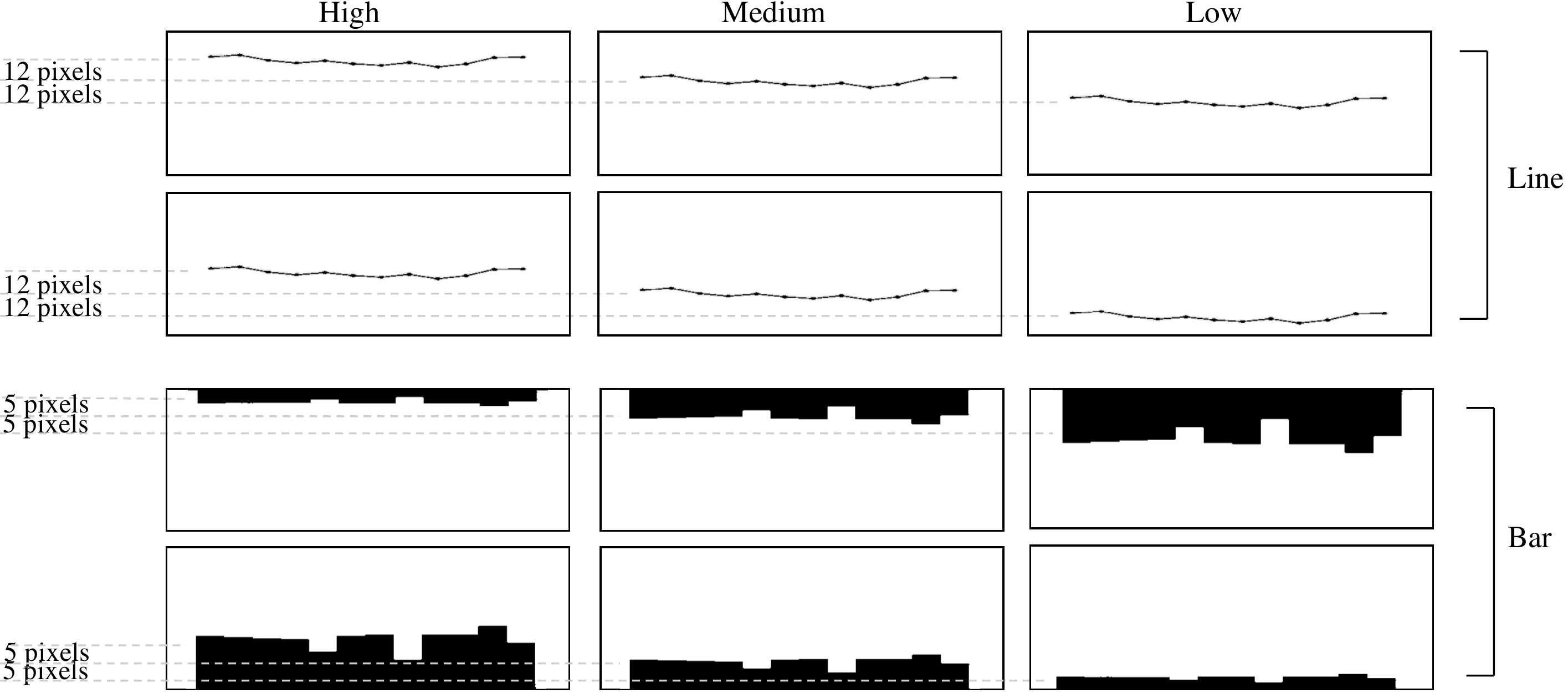}
 \caption{Mean values generated for a line or a set of bars based on the pilot experiment. All stimulus displays were then created around these three set mean values (high, medium and low) for lines and bars located in the top half or bottom half of the display.}
 \label{fig:meanValues}
\end{figure}

In order to pick a set of mean values that would produce the approximately same response precision for lines as for bars, we conducted a two-alternative forced choice pilot experiment with 150 trials each per five participants. Participants were tasked to discriminate between two consecutively graphed data series (two lines or two sets of bars) and report whether the second data series was higher or lower than the first. The two lines or two sets of bars could be 1 to 15 vertical pixels apart. Following conventions from psychophysics, an ideal distance between means would have participants accurately discriminating them in at least 75\% of all trials \cite{green1966signal}, which is roughly halfway between chance (50\%) and perfect (100\%) performance in similar perceptual discrimination tasks. We found that for line displays, participants were at least 75\% accurate in this pilot discrimination task when the three means were separated by 12 pixels (top line accuracy: 93\%; bottom line accuracy: 80\%). For bar displays, participants were at least 75\% accurate when the three means were separated by 5 pixels (top bar accuracy: 93\%; bottom bar accuracy: 87\%). Lines and bars in the display frame were then generated using these mean values (see Figure \ref{fig:meanValues}), with added noise from the normrnd() MATLAB function.

The procedure summary is visualized in Figure~\ref{fig:expTrial} for all three experiments. Participants were cued to attend to one line or one set of bars, also referred to as the ``target line'' or ``target bars'', respectively. They then viewed a display frame with one randomly positioned line, one randomly positioned set of bars, or a combination of lines and/or bars for 500ms, followed by a visual noise mask. Participants responded by dragging a response probe to report the perceived average position of the target on an empty display frame. Depending on the precue, this response probe could be a horizontal line, or a filled rectangular bar (which could be anchored to the top or bottom of the display). The response probe line had a fixed horizontal width and a vertical position that was controlled by the mouse. The response probe bar had a fixed horizontal width and the top of its vertical edge was controlled by the mouse. Changing mouse position would change the height of the rectangular bar. At the beginning of the response, the vertical position of the mouse was randomly set to be at the top of the display or the bottom of the display frame. Response time was unlimited. After participants responded, a fixation cross appeared, indicating the start of the next trial.


\subsection{Experiment Summary}
In three experiments, we empirically test estimation accuracy of average line and bar positions in simple or complex visualizations. Experiment 1 investigates whether systematic biases exist in position estimates for simple graphs, where a single line or a single set of bars is the only present, graphed data series. Experiment 2 and 3 investigate potential positional biases found in complex graphs. Specifically, Experiment 2 explores how average position perception of a graphed data series can be distorted in the presence of an identical type of data series (e.g., single line presented with another line on the display, or a single set of bars presented with another set of bars). Experiment 3 combines graphed data series (e.g., lines and bars on the same display) to examine how the average perception of one type of data series can be distorted by a different type of data series. Figure {\ref{fig:designSpace} illustrates this design space}.


\input{experiment1.tex}

\input{experiment2.tex}
\input{experiment3.tex}

\section{Perceptual Pull Acts Like An Anchoring Effect}
The effect of perceptual pull observed here seems to share an intriguing property with the cognitive-level anchoring effect. In the anchoring effect, the magnitude of a second `baseline' number affects target estimates, even when these baselines are irrelevant to the estimates. As previously mentioned, a population guess for Nova Scotia would be \textit{higher} when first asked if the population is higher or lower than 200,0000 (high magnitude baseline), and \textit{lower} if that question included the number 20,000 (low magnitude baseline). In comparison, perceptual pull is exerted on a target data series when an irrelevant data series acts as a non-target baseline. Across Experiment 1 -- 3, position estimate distributions for a target data series (which remained the same) changed when an irrelevant, non-target data series was introduced. This change depended on the exact position (or magnitude) of that non-target data series. Similar to the cognitive anchoring effect, the observed perceptual pull was greater when the baseline series appeared to be farther away (high magnitude baseline). 

\section{Limitations and Future Directions}
\label{limitation}

\textbf{Asymmetrical Biases} Why do lines and bars show asymmetrical biases? In order to encode average information about lines in a line graph, one must use position. For a set of bars, in contrast, one can use a combination of position, length, or area \cite{yuan:2018:Perceptual} to acquire this information. 
Might that addition of length or area for encoding average bar information be the reason why the average position of bars is overestimated, while there is an underestimation in average line positions?

\textbf{Aspect Ratio} Could the cause of these positional biases be the result of another property of the sets of bars: their aspect ratios? The present work asked participants to estimate the average of a set of bars that had a large width:height aspect ratio across the entire set. While this type of ensemble is commonly seen in visualizations (e.g., a data series displayed as adjacent (thin) bars, showing data over time), it is possible that the wide aspect ratio of these bars contributed to the positional overestimation bias. For example, the representation of the short height of a set of bars could have been intermixed with the representation of its long width, resulting in the bars being remembered as taller and, therefore, leading to consistent overestimation \mbox{\cite{ceja:2019:AspectRatio}}. Would manipulating this aspect ratio lead to changes in the overestimation of the average position of bars within a bar graph? 

\textbf{Figure-Ground Encoding} Attention has been found to be biased toward the ground plane, such that even infants attend to distance-related information provided by objects on the ground compared to objects on the ceiling, even when both are equally visible \mbox{\cite{kavvsek:2013:Ground}}. In our displays, a line may appear to `float' above the x-axis `ground'\footnote{Negative lines that `hang' below the x-axis `ceiling' were not tested.}, while bars `grow' from the `ground' or 'hang' from the `ceiling' in the display frame. One could speculate that lines are systematically underestimated because the ground plane draws attention toward itself. Upward-pointing bars might be overestimated because the `peaks' of the data series draw attention. But why do negative, downward-pointing sets of bars still produce estimates biased toward the top of the screen, instead of downward toward their peaks? 

One possibility is that figure-ground assignment biases found in the visual system \mbox{\cite{wagemans:2012:Gestalt}} may cause viewers to implicitly segment the white negative space under the bars as the `figure'. Overestimating the average of that negative space could lead to an underestimation of the positive space, producing an overestimation of the average position of the downward-pointing bars. 
Future work could test whether some of the present effects found in this current work are driven by how the visual system implicitly segments visualizations into figure-ground.

\textbf{Task Beyond Averaging} What are the root causes of these perceptual biases? Are these roots perceptual and mandatory, such as the mandatory pooling of distributed neural codes for the positions of otherwise separate objects \mbox{\cite{haberman:2012:Ensemble}}? Or might they be more cognitive and strategic, leaving more room for strategy interventions to mitigate the biases? Manipulating the viewer's task could determine whether the bias stems from averaging ensemble positions, or is more generally a bias of perceiving even a single mark position. 

Uncovering the roots of these perceptual biases would be useful to the data visualization community to the degree that it inspires design guidelines for mitigating these effects in real displays. Different underlying mechanisms might predict that the biases would disappear with greater physical separability between graphed data series, separate axes, grid lines, etc.

\textbf{Reporting Mechanisms}  The present study utilized a reporting mechanism in which participants dragged a response probe vertically across the screen to report average position estimates. While a small effect, the initial position of the response probe did account for some amount of the biases observed. We suspect that the response probe may also have had a potential perceptual pull effect on the position estimates. Therefore, future work should test whether the biases found in the current study are also present across alternative reporting mechanisms, such as verbal reports, dragging a dotted line, or comparisons to sequential data series (e.g., is the new data series higher or lower, on average, than the previously shown data series?). 

\textbf{Complex Real-World Stimuli}  Real-world visualizations tend to be more noisy or cluttered in an effort to convey information. Are the position biases found in this current work heightened or diminished in such complex visualizations? Future iterations of this work should examine this position perception bias within real-world visualizations that are not as simple as those used in the current experiments. For example, how would the effect of perceptual pull fare in complex line graphs with multiple lines of different colors (as commonly seen in an analytics dashboard)?

\textbf{Untested Encodings}  Finally, reports of average position can be biased, but do these biases persist in memory for values encoded in orientation, size, saturation, or other forms of data encodings? The perceptual psychology studies reviewed in the introduction section suggest that all of these data encodings can lead to systematic biases, even using displays that are quite similar to those in data visualization research. Could these values be underestimated or overestimated in memory, similar to position? 

\section{General Conclusion and Design Guidelines}
In three experiments, we empirically tested how average position reports for a single graphed data series (a line or a set of bars) can be biased. We found that reports \textit{underestimated} average line positions and \textit{overestimated} average bar positions. In displays containing two graphed data series (lines and/or sets of bars), we observed an effect of ``perceptual pull'', in which an \textit{irrelevant} data series `pulled' average positional estimations of the \textit{relevant} data series in its direction. 

At this early stage of research, we hesitate to provide firm design guidelines. But, because position estimation biases were smaller and judgments were more precise for bars compared to lines, a visualization designer might consider using bars to display their data, in the absence of other constraints. Designers seeking the confidence of avoiding perceptual pull effects could avoid plotting two series in the same display. After answering these questions listed in Section 10, the collective view of the literature should provide stronger guidance on how to predict and avoid biases for online average estimates (especially within a dual y-axis design), comparisons of online averages to averages from memory, or comparisons of two averages from memory.

\bibliographystyle{abbrv}
\bibliography{template}

\end{document}

%% file: experiment1.tex
\section{Experiment 1}
Experiment 1 tests how accurately people can perceive average positions of a single line or single set of bars in a graph. This experiment establishes a baseline to understand how a \textit{single} graphed data series is perceived without the potential influence of other graphed data series. These findings will be later compared to how the presence of an \textit{additional} graphed data series may further bias perception. 

\subsection{Design and Procedure}
This experiment investigates whether people report average line and bar positions in graphs in a biased way by comparing participants' \textit{estimation} of average line or bar positions to the \textit{true average position} of the line or bars in a mixed-model design. 


In Experiment 1, participants were cued before each trial with the task of either estimating the average position of a line or a set of bars to be shown on a subsequent stimulus display (see Figure~\ref{fig:expTrial}). Depending on that precue, participants then saw a stimulus display that contained a set of bars or a line appearing on the top or bottom half of the display (see Figure~\ref{fig:expTrial}). As shown in Figure~\ref{fig:designSpace}, these data series could be uniform (where all points on the line or on the set of bars are of the same value), or noisy (where all points are of different values). 

For a total of 576 trials, each participant completed 288 trials for each line and bar position estimate, with half of trials for each condition displaying a noisy version of the data series and the remaining half of the trials displaying a uniform data series. For analysis, we examined the average position estimations participants made across all these dimensions. We also examined the effect of the initial location of the response probe.

Thirteen undergraduate students from Northwestern University ($M_{age}$=18.62 years, $SD_{age}$=0.65) participated in exchange for course credit in an introductory psychology class. We excluded one participant who did not complete the experiment from our data analysis.

\subsection{Underestimation of Lines}
We used a mixed-effect linear model to predict estimated position with fixed effects of the actual display location of the line (i.e., the pixel value and whether it was on the top or bottom half of the display), whether the line was noisy or uniform, the initial location of the response probe, practice effect (as trial number), and participants as random intercept. In this model, we excluded all trials in which participants made an obviously wrong estimate, defined as making an estimate in the \textit{bottom quarter} of the display for a stimulus appearing in the \textit{top half} of the display, or vice versa. 

\begin{figure*}[t!]
 \includegraphics[width=18.1cm]{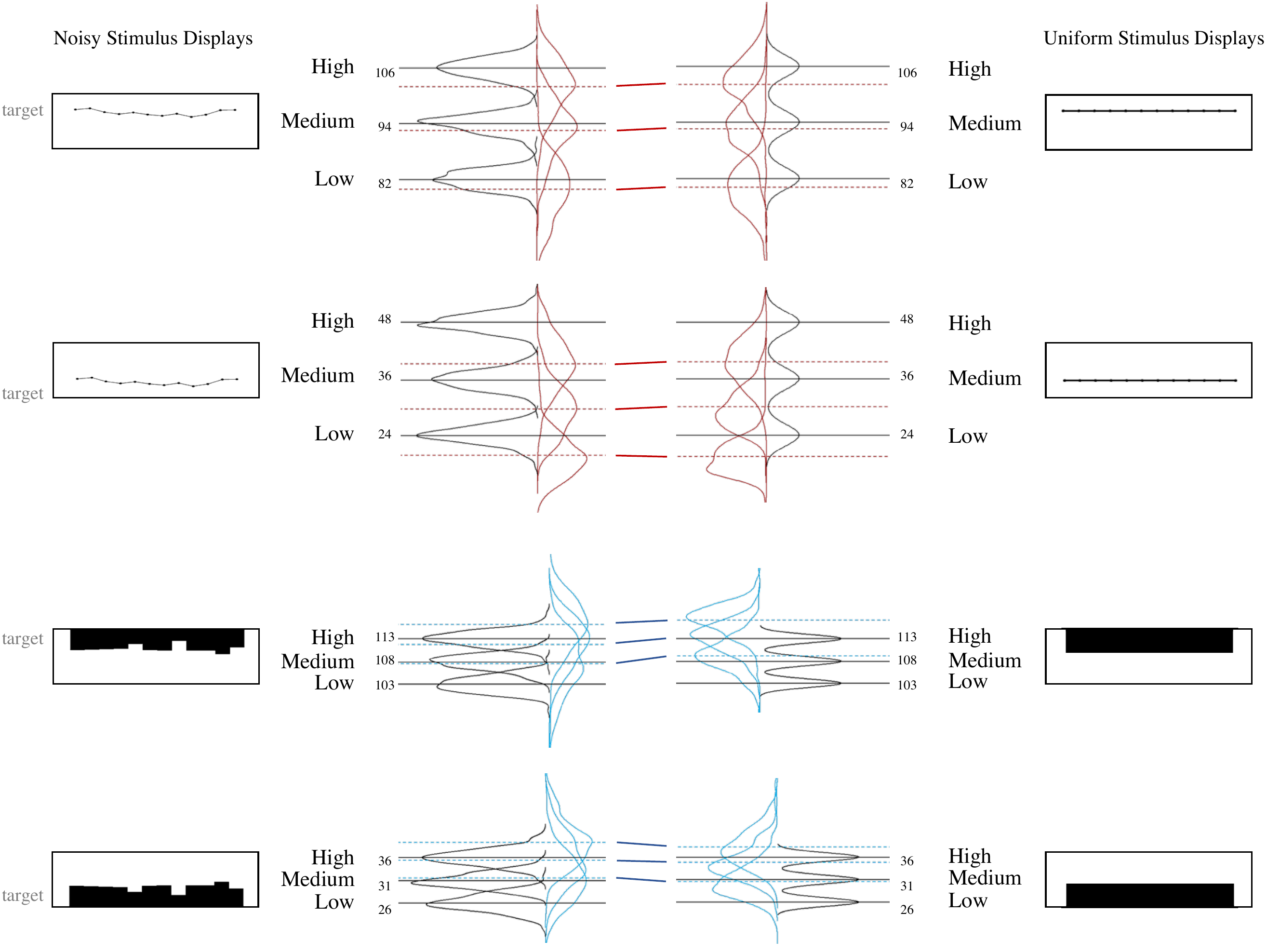}
 \caption{Results from Experiment 1 for noisy and uniform simple graphs. First and fourth columns: Example displays representing the 'medium' mean position for noisy and uniform displays, respectively. Second and third columns: Density curves for true average position distributions \textit{(solid black lines for each of the three means: high, medium, low; pixel values given)}, and density curves for the estimated average position distributions for each of the three means \textit{(line estimates: solid red lines; bar estimates: solid blue lines)}. The orientations of the solid red or blue lines (shown between the density curves for the estimated average position distributions) show the differences in position estimates between noisy and uniform conditions.}
 \label{fig:exp1Density}
\end{figure*}

As shown in Figure~\ref{fig:exp1Density} (top two rows), we observed an overall underestimation of line position, in which participants estimated average line positions to be lower than where the average position of the line actually appeared ($MD_{overall}$=-4.49 pixels, $SE$=0.11, $Est$=1.01, $CI_{95\%}$=[0.95, 1.07], $p$<0.001, $\eta_{partial}$=0.68). This underestimation persisted regardless of whether the line appeared on the top or bottom half of the display screen ($MD_{bottom}$=-6.50, $SE_{bottom}$=0.15, $MD_{top}$=-2.47 pixels, $SE_{top}$= 0.16, $Est$=-5.57, $CI_{95\%}$=[-8.45, -2.70], $p$<0.01, $\eta_{partial}$=0.96), although participants underestimated the average of the bottom line more than the average of the top line. Estimation error did not depend on whether the line was noisy or uniform ($Est$=0.12, $CI_{95\%}$=[-0.28,0.53], $t$=0.61, $p$=0.54, $\eta_{partial}$=0.00). This provides further evidence that the underestimation of average line positions is not an artifact of the noise in the line stimulus, as underestimation occurred for even uniform lines on the display. 

There was also a very small interaction between the average pixel position of the line and whether that line was presented in the top half or bottom half of the display ($Est$=-0.094, $CI_{95\%}$=[-0.13, -0.054], $p$<0.001, $\eta_{partial}$=0.006). For lines appearing in the top half of the display, participants underestimated average position \textit{less} when the lines were located closer to the center of the screen. For lines appearing in the bottom half of the display, participants underestimated average position \textit{less} when the lines were located closer to bottom boundary of the display box (further away from center). 

\begin{figure}[t!]
 \centering
 \includegraphics[width=8cm]{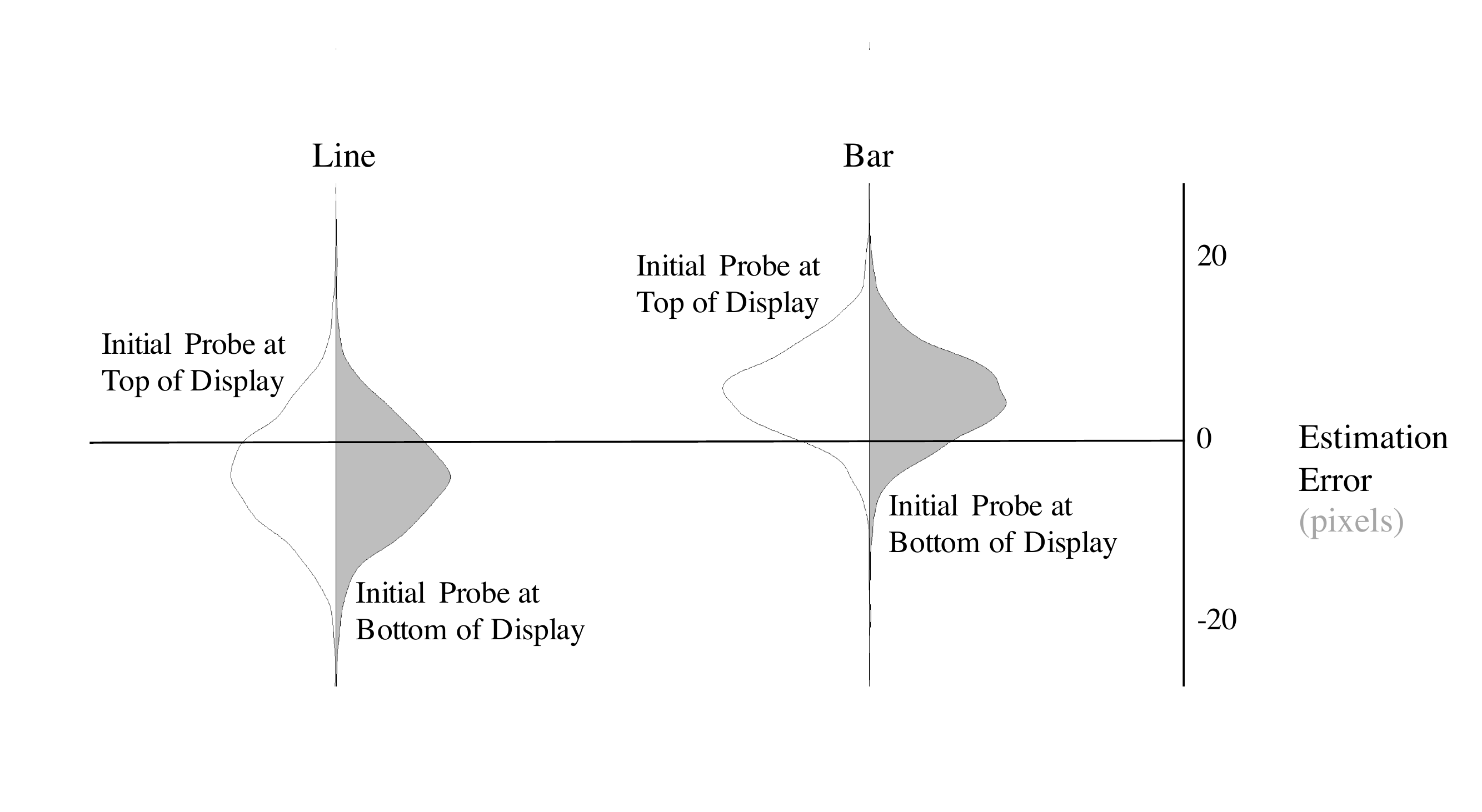}
 \caption{Difference between the initial probe location in average line and bar estimation tasks.}
 \label{fig:probe}
\end{figure}

We found no practice effect ($\eta_{partial}$=0.00, $p$=0.45), suggesting that average position estimations did not get more or less biased as participants completed more trials. We also found that while initial probe location (where the response probe for position estimations was initially presented in the display) had a significant influence on the estimation error ($Est$=0.62, $CI_{95\%}$=[0.21, 1.02], $p$=0.0022, $\eta_{partial}$=0.003), its effect size was small (see Figure~\ref{fig:probe}).

Was this systematic underestimation an artifact of poor average estimation strategies? We considered whether the underestimation in average line position was the result of participants simply choosing the lowest point on the noisy line as their response for the average position of the line. To test this, we compared participants' estimated average line positions with the lowest point on the noisy line using a different mixed-effect linear model, with estimated line locations as the dependent variable (DV) and the position of the lowest point on the noisy line as the independent variable (IV). If participants estimated the average line only relying on the position of the lowest point, the slope of the linear model should be 1. Using a Wald test with confidence intervals at 95\%, we found that the test slope of the mixed-effect model was in the range of [0.64, 0.70] for lines on the top of the display, and in the range of [0.57, 0.63] for lines on the bottom of the display. Neither range included the value 1, which suggests that participants did not base their average position estimations only on the lowest point for lines on both the top and bottom half of the display. We did a similar comparison predicting whether estimated average positions depended on the average of the highest and lowest points on the line, and found the slope of this regression line to also not include 1 in its 95\% confidence interval, ($Est_{top}$=[-0.057, 0.12], $Est_{bottom}$=[-0.10, 0.06]). This suggests that the participants were not simply averaging the highest and lowest points on the line stimulus to make their estimations.

\begin{figure*}[t]
 \centering
 \includegraphics[width=18.1cm]{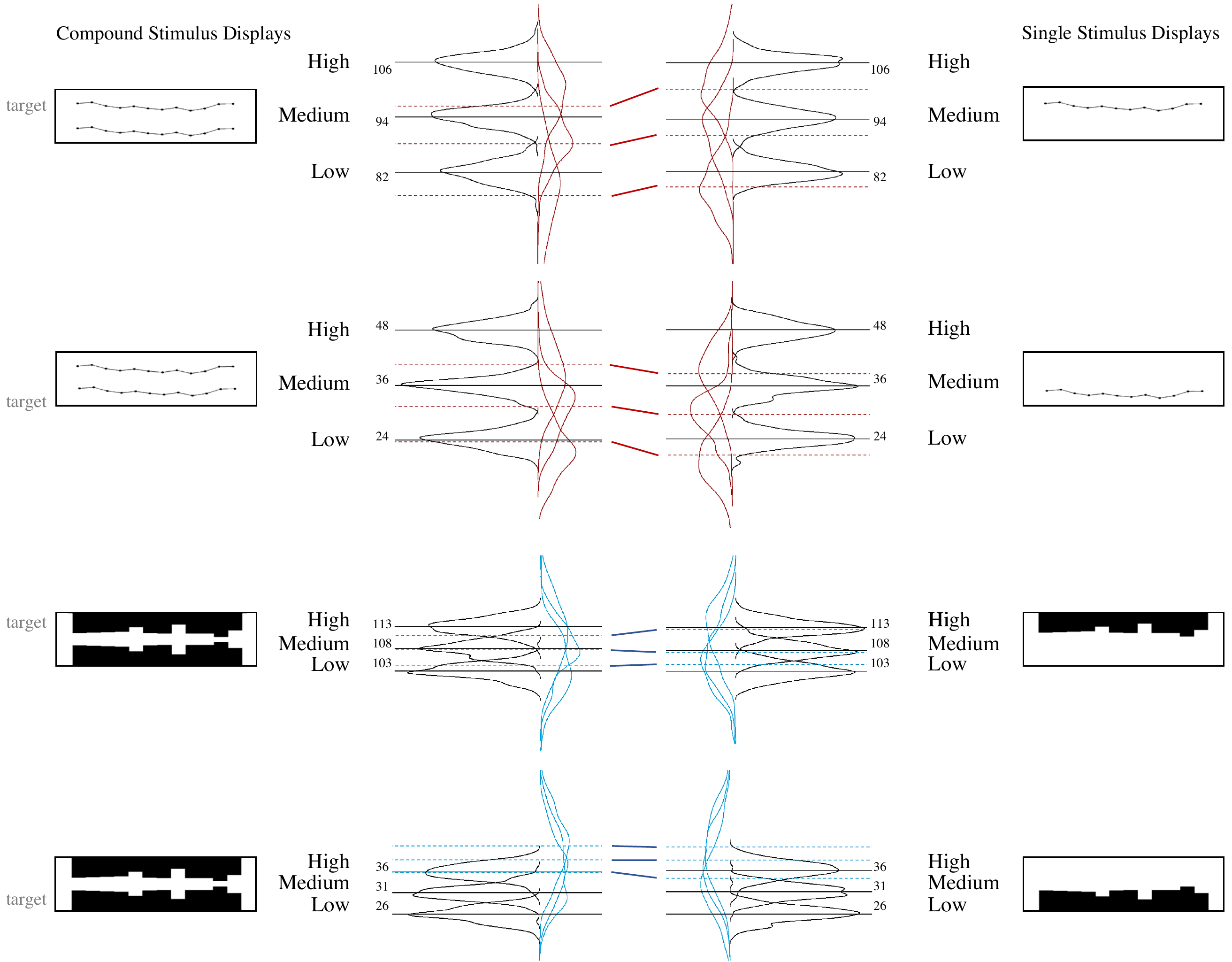}
 \caption{Results from Experiment 2 for compound stimulus displays (complex graphs; i.e., two lines or two sets of bars) and single stimulus displays (simple graphs; i.e., one line or one set of bars). There is still consistent underestimation of lines and overestimation of bars, but these effects are exaggerated or diminished by perceptual pull, depending on the location of the graphed data series. The orientations of the solid red or blue lines (shown between the density curves for the estimated average position distributions) show the differences in estimates between compound and single stimulus displays.}
 \label{fig:exp2Results}
\end{figure*}

\subsection{Overestimation of Bars}
We utilized the same mixed-effect linear model to examine estimated average bar positions. As shown in Figure~\ref{fig:exp1Density} (bottom two rows), we observed an overall overestimation of bar position, where participants systematically estimated average bar positions to be higher than its actual position ($MD_{overall}$=4.19 pixels, $SE$=0.086, $Est$=0.72, $CI_{95\%}$=[0.59, 0.84], $\eta_{partial}$=0.33, $p$<0.001). This overestimation persisted regardless of whether the bar stimulus appeared on the top ($MD_{top}$=4.43, $SE_{top}$=0.12) or bottom half of the display screen ($MD_{bottom}$=3.95, $SE_{bottom}$=0.12, $Est$=10.26, $CI_{95\%}$=[4.00, 16.51], $p$<0.001, $\eta_{partial}$=0.98), although participants overestimated the average position of the bottom bars significantly more than the average position of the top bars. There was no effect of whether the bars were noisy or uniform ($MD$=0.15, $SE$=1.33, $Est$=0.12, $CI_{95\%}$=[-0.21, 0.45], $p$=0.57, $\eta_{partial}$=0.00), suggesting that this systematic bias may not be the result of participants utilizing outliers for position averages. There was no practice effect ($\eta_{partial}$=0.00, $p$=0.85), but an effect of initial probe location ($\eta_{partial}^2$=0.007, $p$<0.001) with a small effect size (see Figure~\ref{fig:probe}). 

We then further investigated whether this overestimation was an artifact of participants basing average bar position estimations on the highest point on the bar graph. We compared participants' estimated average bar positions with the highest point on the noisy set of bars using the same method as before, and found the 95\% confidence interval for the test slope of the model to be in the range of [0.29, 0.36] for the bottom set of bars and [0.22, 0.29] for the top set of bars. Since 1 is not included in either confidence interval, the results suggest that participants did not simply base their average position estimations on the highest point on the bars. Similar comparisons between estimated average position and the average of the highest and lowest points on the bars also found that the slope of the regression line did not include 1 in its 95\% confidence interval ($Est_{top}$= [0.01, 0.11], $Est_{bottom}$=[0.22, 0.29]), suggesting that the participants were not simply averaging the highest and lowest points on the set of bars to generate their average positional estimates.

\subsection{Discussion}
Experiment 1 illustrated a systematic underestimation of average line positions and overestimation of average bar positions. Interestingly, this effect occurred regardless of whether the lines and bars were noisy or uniform. Comparing the variance of the estimations via an F-test, we found that while participants estimated uniform lines and uniform sets of bars with more precision and less variance than with noisy lines and noisy sets of bars ($F(1,3454)$=1.1479, $p$<0.0001), biases were still prevalent in both noisy and uniform displays. Overall, this experiment showcased that position encoding is not immune to perceptual bias.

%% file: experiment2.tex
\begin{figure*}[ht]
 \includegraphics[width=18.1cm]{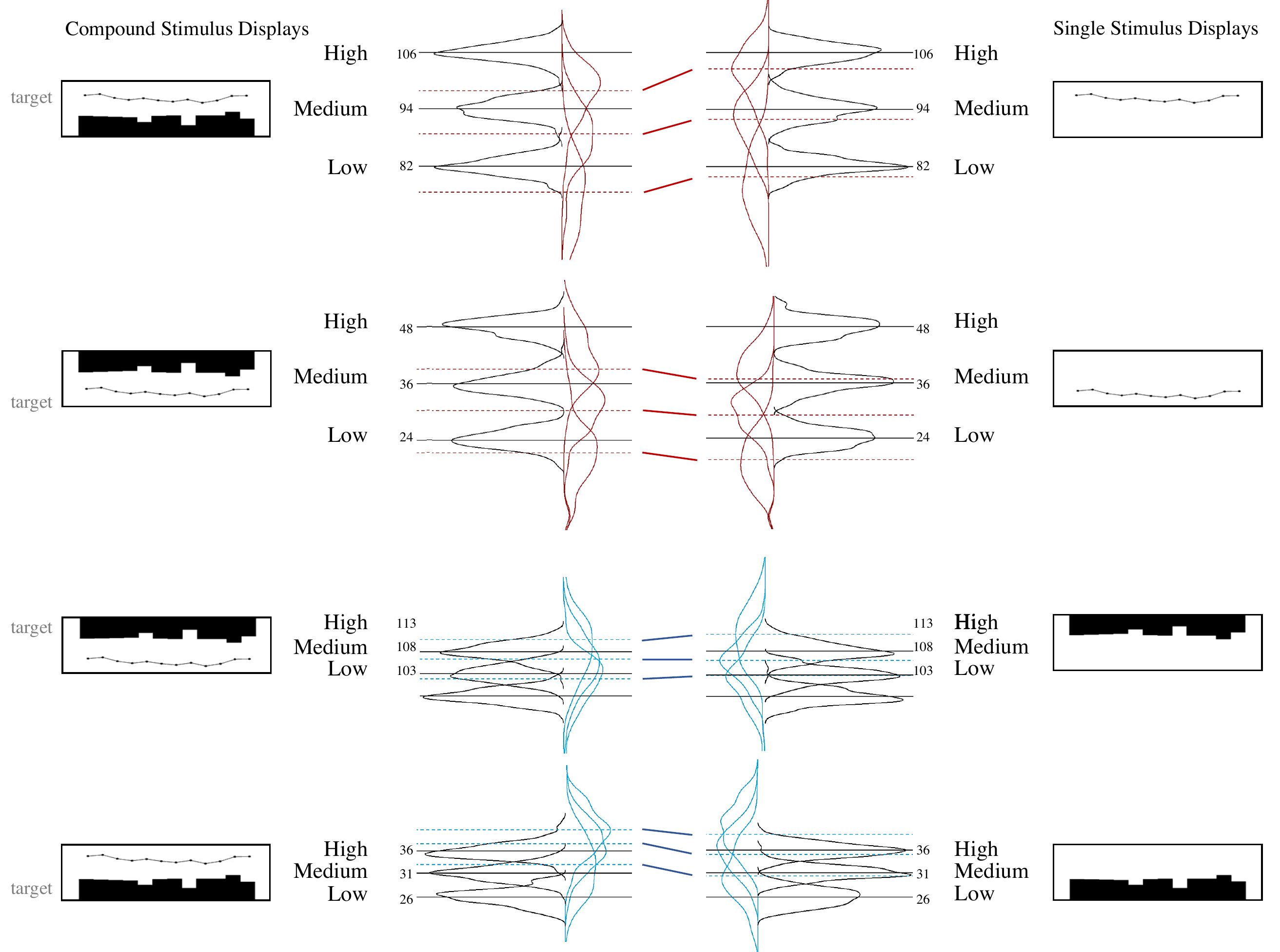}
 \caption{Results from Experiment 3 for compound stimulus displays (i.e., one line and one set of bars) and single stimulus displays (i.e., one line or one set of bars). Perceptual pull does not depend on the characteristics of the other graphed data series.}
 \label{fig:exp3Results}
\end{figure*}

\section{Experiment 2}
Experiment 1 provided evidence for biased reports of average position for a simple graph with one data series, but how is this bias affected by the presence of an additional data series within a complex graph? Experiment 2 expands on this investigation of biases by determining whether the perception of average line and bar positions can be further biased by other lines and bars present on the same graph, respectively. We first explore how the position of one line may influence the perceived position of a target line within a display containing two lines (referred to as ``compound line-line'' displays). We then test the effect of an additional set of bars on the average position estimations of a target set of bars in a display with two sets of bars (referred to as ``compound bar-bar'' displays). 

\subsection{Design and Procedure}
The methods were similar to those in Experiment 1, with the following changes. Participants were presented with noisy compound line-line or noisy compound bar-bar displays and were precued to report the average position of a line or a set of bars presented at the top or bottom of the display (see Figure~\ref{fig:expTrial}). 

For a total of 240 trials, each participant completed 120 trials for each line and bar average position estimation condition. For half of these trials, participants were tasked with judging the position of the top data series and in the other half, the bottom data series. We also included 144 control trials in which each participant estimated the average position of a single noisy line or a single noisy set of bars (referred to as ``single-line'' or ``single-bar'' displays; see Figure~\ref{fig:designSpace}), replicating Experiment 1.

Twelve different undergraduate students from Northwestern University ($M_{age}$=19.31 years, $SD_{age}$=1.55) participated in exchange for course credit in an introductory psychology class. 

\subsection{Underestimation and Overestimation}
Using the same analysis method as Experiment 1, Experiment 2 replicated the results of Experiment 1 in the single-line and single-bar conditions. 

In the single-line displays, participants still underestimated average line positions ($MD$=-5.27, $SE$=0.18, $p$<0.001) (see first two rows in Figure~\ref{fig:exp2Results}). Further analyses showed no practice effect ($\eta_{partial}^2$=0.00, $p$=0.91) and no effect of initial location of the response probe ($\eta_{partial}^2$=0.003, $p$=0.12).

In single-bar displays, participants overestimated average bar positions ($MD$=3.39, $SE$=0.17, $p$<0.001) (see last two rows in Figure~\ref{fig:exp2Results}). We found no practice effect ($\eta_{partial}^2$=0.001, $p$=0.23), but a small effect of the initial response probe location ($\eta_{partial}^2$=0.004, $p$=0.0025).

\subsection{Perceptual Pull}
For compound line-line displays and compound bar-bar displays, we observed an underestimation in average line positions and an overestimation in average bar positions. Additionally, we also found an effect of ``perceptual pull'': position estimates for a target data series (a set of bars or a line) were `pulled' toward the irrelevant data series shown on the same graph.

We used another mixed-effect model predicting estimation error with fixed effects of whether the data series was present in the top half of the display or the bottom half, whether the display was compound or single, and trial number, and a random effect of participants. In compound line-line displays, there were no main effects of line location (top or bottom) ($Est$=-2.05, $CI_{95\%}$=[-2.22, -0.44], $p$=0.82, $\eta_{partial}$=0.00) or display type (compound or single) ($Est$=-2.05, $CI_{95\%}$=[-3.07, -1.02],$p$=0.82, $\eta_{partial}$=0.00). However, there was a significant interaction between line location and display type, such that the magnitude of line position underestimation between lines that appeared on the top and bottom half of the display differed ($Est$=3.75, $CI_{95\%}$=[2.30, 5.19], $p$<0.001, $\eta_{partial}$=0.011). Underestimation of the top line was \textit{exaggerated}, such that participants underestimated the top line even more compared to single-line displays ($MD_{top,single-compound}$=-1.92, $SE$=0.71). In contrast, underestimation of the bottom line was \textit{reduced} compared to single-line displays ($MD_{bottom,single-compound}$=2.01, $SE$=0.65). 

From this interaction between display type and location of the line (top or bottom) on position estimation, we speculate that the top and bottom lines `pulled' position perception toward one another. The presence of the irrelevant top line further biased perception of the average position of the bottom target line, and the presence of the irrelevant bottom line further influenced perception of the average position of the top target line. 

We also observed the same perceptual pull in the compound bar-bar displays using an identical mixed model. There was a significant main effect of bar position such that bottom bars were significantly more overestimated than top bars ($Est$=-8.08, $CI_{95\%}$=[-8.81, -7.34], $p$<0.01, $\eta_{partial}$=0.22). There was no main effect of display type ($Est$=-0.61, $CI_{95\%}$=[-1.46, 0.23], $p$=0.63, $\eta_{partial}$=0.00), but there was a significant interaction between bar location and display type ($Est$=1.51, $CI_{95\%}$=[0.32, 2.71], $p$=0.013, $\eta_{partial}$=0.003). Overestimation of the top bars was \textit{reduced}\footnote{Note that this is a decrease in absolute position in the reference frame of the display, meaning participants perceived the top set of bars as vertically longer in the compound bar-bar displays than the single displays.}, such that participants overestimated the top bars less compared to single-bar displays ($MD_{top, single-compound}$=0.97, $SE$=0.44). In contrast, overestimation of the bottom set of bars was \textit{exaggerated} more compared to single-bar displays ($MD_{bottom, single-compound}$=-0.53, $SE$=0.48). This interaction again suggests that, similar to the lines, there exists a perceptual pull between bars, `pulling' the perception of their average positions toward one other.

\subsection{Discussion}
Experiment 2 replicated evidence that perception of average line and bar positions is biased. Furthermore, it showcased a perceptual pull effect, in which the presence of an irrelevant line or set of bars in the same display pulled average position estimations of a target line or set of bars toward the position of this irrelevant data series (see Figure~\ref{fig:exp2Results}).

%% file: experiment3.tex
\section{Experiment 3}
Experiment 2 illustrated the existence of perceptual pull, but what determines the extent of the perceptual pull? Is it data series-dependent (in which a specific type of graphed data series could influence the strength of the perceptual pull more than another type)? Is perceptual pull dependent on sufficient perceptual similarity between the two graph elements (i.e., irrelevant bars would exert a larger influence on average bar judgments than on average line judgments)? Experiment 3 expands upon Experiment 2 by diversifying the types of graphed data series present in the display to test the extent to which a line and a set of bars on the same graph can influence each other.

\subsection{Design and Procedure}
The methods used in Experiment 3 were similar to that of Experiment 2. We will refer to displays with a line present in the top half and a set of bars in the bottom half of the display as ``compound line-bar'' displays, and displays with a set of bars on the top half and a line on the bottom half of the display as ``compound bar-line'' displays (see Figure~\ref{fig:designSpace}).

Twelve different undergraduate students from Northwestern University ($M_{age}$=19.00 years, $SD_{age}$=1.04) participated in this experiment in exchange for course credit in an introductory psychology class. 

\subsection{Underestimation and Overestimation}
Experiment 1 and 2 results were replicated. 

In single-line displays, there was still a systematic underestimation of average line positions as lower than their true positions ($MD$=-6.47, $SE$=0.26, $p$<0.001), and an overestimation of average bar positions as higher than their true positions ($MD$=3.85, $SE$=0.19, $p$<0.001) (see far right of Figure~\ref{fig:exp3Results}). There was no evidence of a practice effect in either the line-bar ($p$=0.36, $\eta_{partial}$=0.00) or the bar-line ($p$=0.23, $\eta_{partial}$=0.00) displays, but there was an effect of the initial location of the response probe with a small effect size for both line-bar ($p$<0.001, $\eta_{partial}^2$=0.03) and  bar-line conditions ($p$<0.001, $\eta_{partial}^2$=0.02).

\subsection{Perceptual Pull}
As in Experiment 2, we still found an underestimation in average line positions and an overestimation in average bar positions for all compound displays. Similarly, we observed an effect of perceptual pull between the two data series (one line and one set of bars) in the display.

We used a mixed-effect model with display type (single or compound), graphed data series type (line or bar), the interaction between display type and graphed data series type, trial, and initial probe position as fixed effects, and with participants as a random effect. In the compound line-bar condition, we found a significant main effect of graphed data series type ($Est$=-11.14, $CI_{95\%}$=[-12.16, -10.12], $p$<0.001, $\eta_{partial}$=0.188), a significant interaction of display type and graphed data series type ($Est$=4.72, $CI_{95\%}$=[3.06, 6.39], $p$<0.001, $\eta_{partial}$=0.014), a small effect of initial probe location ($Est$=3.41, $CI_{95\%}$=[2.60, 4.21], $p$<0.001, $\eta_{partial}$=0.188), and negligible effects from other predictors (all $\eta_{partial}$<X, all $p$s>X). The underestimation of the top line was \textit{exaggerated} when compared to the single-line displays ($MD_{topLine, single-compound}$=-2.96, $SE$=0.90, $p$<0.001), as the bottom bars `pulled' the average positional percept of the top line down. Overestimation of the bottom set of bars was also \textit{exaggerated}, such that participants overestimated the bar position more when there was an above line `pulling' the average positional percept of the set of bars up ($MD_{bottomBar, single-compound}$=-1.37, $SE$=0.53, $p$<0.001).

Similarly, in the compound bar-line condition, a mixed-effect model showed a significant main effect of graphed data series type ($Est$=7.90, $CI_{95\%}$=[6.94, 8.85], $p$<0.001, $\eta_{partial}$=0.209), a significant interaction effect between display type and graphed data series type ($Est$=3.45, $CI_{95\%}$=[1.88, 5.01], $p$<0.001, $\eta_{partial}$=0.008), a small effect of initial probe location ($Est$=2.86, $CI_{95\%}$=[2.10, 3.62], $p$<0.001, $\eta_{partial}$=0.023), and negligible effects from other predictors (all $\eta_{partial}$<0.01, all $p$s>0.10). Overestimation of the top set of bars was significantly \textit{reduced} when a line appeared below its position on the display ($MD_{topBar, single-compound}$=1.49, $SE$=0.67, $p$<0.001), suggesting that the bottom line `pulled' the average positional percept of the set of bars down. Underestimation for the bottom line was also \textit{reduced}, such that the average position for the bottom line was underestimated less compared to single displays ($MD_{bottomLine, single-compound}$=1.50, $SE$=0.65, p<0.001), as the top set of bars `pulled' the average positional percept of the bottom line up. 

In the compound displays overall, the effect of perceptual pull \textit{exaggerated} the underestimation of average line positions and the overestimation of average bar positions when a line was located in the top half and a bar was located in the bottom half of the display. But perceptual pull \textit{reduced} the same underestimation and overestimation bias when a line was located in the bottom half and a bar was located in the top half of the display.

\subsection{Strength of Influence}
The effect of perceptual pull occurs across graphed data series type, but is the \textit{extent} of this perceptual pull dependent on the type of data series present? In other words, would a data series pull the same or different type of series more strongly, or would it pull all data series equally? 

\begin{figure}[t]
 \includegraphics[width=8.8cm]{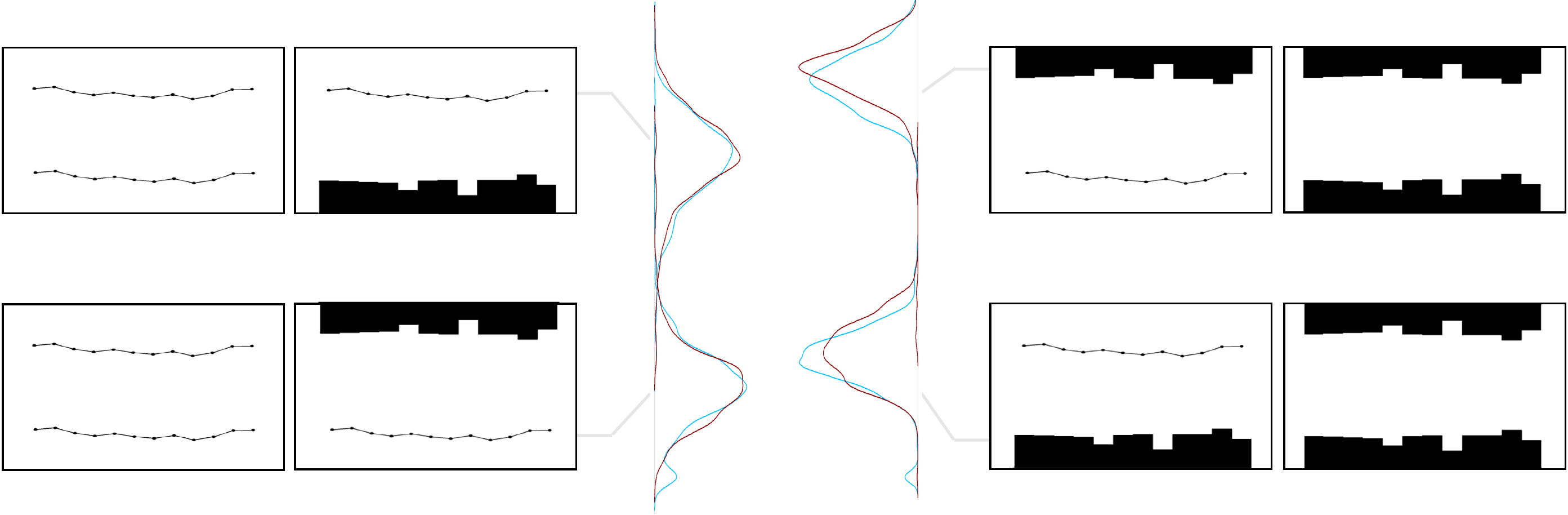}
 \caption{Estimation error for target data series is not dependent on the irrelevant graphed data series. Left three columns: Comparing average estimated line positions when the non-target data series is a line (\textit{red}) or a set of bars (\textit{blue}). Right three columns: Comparing estimated bar positions when the non-target data series is a line (\textit{red}) or a set of bars (\textit{blue}).}
 \label{fig:nonTargetInfluence}
\end{figure}

With data from Experiment 2 and 3, we conducted a between-subject ANOVA comparison examining the variation in average line and bar position estimations depending on the non-target data series (a line or a set of bars) (see Figure~\ref{fig:nonTargetInfluence}). Neither top nor bottom target line position estimations in a compound line-line display were significantly different from that in a compound line-bar display (top: $MD$=0.46, $SE$=0.70, $p$=0.57, with Tukey correction; bottom: $MD$=1.31, $SE$=0.56, $p$=0.42). Similarly, neither top nor bottom target bar position estimations in a compound bar-bar display were significantly different from that in compound bar-line displays (top: $MD$=3.11, $SE$=0.52, $p$=0.21; bottom: $MD$=1.98, $SE$=0.43, $p$=0.29). This suggests that the extent of perceptual pull does \textit{not} depend on data series type.
 
\subsection{Discussion}
Experiment 3 showed that perceptual pull is not dependent on graphed data series type, but can generalize across data series (e.g., lines and/or bars). A single, irrelevant line has a similar pulling force for a target set of bars as for a target line, and vice versa for a single set of irrelevant bars on a target set of bars or a target line. 